\documentclass[11pt,twocolumn]{article}
\usepackage[dvips]{graphicx}
\usepackage{vmargin}  
    \setpapersize{A4}
    \setmarginsrb{1.3cm}{2cm}{2cm}{2.5cm}{0pt}{0mm}{0pt}{0mm}
\usepackage{newcent}  
\usepackage[authoryear]{natbib}
     \bibpunct{[}{]}{;}{a}{,}{,}
\usepackage{amsmath}

\begin{document}

\title{Robustness of Transcriptional Regulation\\ in Yeast-like Model Boolean Networks\\}
\author{Murat Tu\u{g}rul$^{1a}$\footnote{corresponding author: mtugrul@ifisc.uib-csic.es and his present address: IFISC(Institute for Cross-Disciplinary Physics and Complex Systems), UIB-CSIC,
 Campus Universitat de les Illes Balears, E-07122 Palma de Mallorca, Spain}, Alkan Kabak\c{c}{\i}o\u{g}lu$^{1b}$\\
 $^{1a}$Computational Sci.\& Eng. Master Programme , $^b$ Physics Department, \\
 Ko\c c University, Sar{\i}yer 34450 Istanbul, Turkey}

\date{\today}

\maketitle

\begin{abstract}
  We investigate the dynamical properties of the transcriptional regulation of
  gene expression in the yeast \textit{Saccharomyces Cerevisiae} within the
  framework of a synchronously and deterministically updated Boolean network
  model. By means of a dynamically determinant subnetwork, we explore the
  robustness of transcriptional regulation as a function of the type of
  Boolean functions used in the model that mimic the influence of regulating
  agents on the transcription level of a gene. We compare the results obtained
  for the actual yeast network with those from two different model networks,
  one with similar in-degree distribution as the yeast and random otherwise,
  and another due to Balcan \textit{et al.}, where the global topology of the yeast
  network is reproduced faithfully. We, surprisingly, find that the first set
  of model networks better reproduce the results found with the actual yeast
  network, even though the Balcan \textit{et al.} model networks are
  structurally more similar to that of yeast.
\end{abstract}

\paragraph{INTRODUCTION}

Recent advances in biotechnology allowed the accumulation of a vast amount of
experimental data on intra-cellular processes, however, our knowledge on how a
cell works remains
incomplete~\citep{Lockhart_Winzeler_Genomics,Barabasi_Oltvai_NetworkBiology}. The
key component of the functional organization in a cell is the regulation of
gene expression. By now, interacting gene pairs for several organisms have
been identified with significant coverage~\citep{Bergman_etal_SixGRN}.
In particular the set of regulatory interactions identified in
\textit{Saccharomyces Cerevisiae} are believed to be close to complete~\citep{Yeastract}.

The activation/suppression dynamics in a cell evolves on a complex and
inhomogeneous network of interactions. Therefore, the framework of graph
theory serve as a powerful mathematical tool for
studying the regulation of the gene expression on cellular
level~\citep{AlbertBarabasi_StatMecCompNets, Bollobas_ModernGraph,
  Milo_etal_Motifs, RichClubCoeffic_Colizza_etal,
  Newman_ScientficCollaborationNetworks_2_2001, Barabasi_Oltvai_NetworkBiology}. The
topology of the graph describing the gene regulatory network (GRN) is far from
being random and has been studied for several organisms, in particular the
budding yeast~\citep{Guelzim_Yeast, Nicholas_etal_GenomicsRegNetworks,
  Bergman_etal_SixGRN}.

Deterministically and synchronously updated Boolean networks have been used
widely as a model for regulatory dynamics~\citep{Kauffman_Network,Aldana_BooleanNetPLtopology2003,Balcan_Erzan2007,BalcanErzan2006}. In this model, the expression levels of
genes are discretized to take values 0 or 1 at each time step. Although it is a
major oversimplification~\citep{Norrell_CompareBooleanAndContinous}, this
approach has proven valuable in the context of gene
regulation~\citep{Mendoza_etal_Arabidopsis, Espinosa_AThalianaFlowerDevelopment,
  AlbertOthemer_TopologyPredictsExpression}.

The network topology of the yeast's GRN is now believed to be unveiled to a
large extent. However the nature of interactions, i.e., the rules that govern
the dynamics, are not known in comparable detail. Accordingly, a statistical
approach involving randomly assigned functions is relevant. Several classes of
such functions have been investigated in the literature. The unbiased choice
is to pick random Boolean functions. On the other hand, it has been claimed
that experimental data is consistent with a subset of Boolean functions where
one of the output is fixed for a particular value of one of the inputs
(canalizing functions)~\citep{Harris_Functions}.  It has also been suggested
that a subset of canalizing functions (nested canalizing functions) is more
appropriate for gene regulation dynamics on yeast~\citep{Kauffman_Nested}. A
more recent study finds that two subclasses of the nested canalizing functions
are actually dominant in the yeast~\citep{Nikolajewa_SpecTypeRuletable}.

The computational bottleneck in the analysis of Boolean network dynamics is
the fact of that number of states increases exponentially with system size.
This makes an exhaustive enumeration prohibitive, even if, in most cases, a
fraction of the nodes can be left outside the analysis due to their
irrelevance to the dynamics by virtue of either the topology or the choice of
the function set~\citep{Socolar_Kauffman}. In this paper, we determine and use a
strongly connected subset of the genes that dictates the network's dynamical
character and use a statistical approach to identify its robustness.

The paper is organized as follows. In the Method section, we present the
yeast's gene regulation network, describe the employed Boolean dynamics,
define the function classes used for setting the rules of the dynamics, propose a dynamically
relevant subnetwork and the
model networks used for comparison with yeast. In the Result section, we present and analysis of the
yeast's GRN dynamics, in particular exploring the robustness of the network
under small perturbations, comparing the results for different types of
functions and for the actual vs. model network topologies. We discuss our findings in the last section.

\paragraph{METHOD \& MODELS}

Transcriptional regulation of gene expression in a cell operates through
transcription factors (TFs). These proteins bind the DNA on ``promoter
regions'' (PRs) that act as the regulation centers of each gene. The details
of this interaction can be very complex. In our study, as in past studies in
the literature, we assume that effect of the TFs that regulate a certain gene
can be summarized in a Boolean function whose inputs represent the presence
or the absence of TFs and the output determines whether the gene is
activated or inhibited for the given expression profile of the TF genes.

The regulation dynamics evolves on a directed graph, whose nodes are the genes
and a directed edge from $A$ to $B$ indicates that the product of $A$
regulates $B$. The corresponding network for {\it Saccharomyces Cerevisiae}
can be retrieved from YEASTRACT database~\citep{Yeastract} (www.yeastract.com).
In order to be able to compare our results with past studies, we here consider
an earlier version (2005) of the network including 4252 genes (with 146 TFs)
with 12541 interactions. As explained below, we also consider two model
networks, one with a similar in-degree distribution as the yeast network
above and random otherwise, and another with a topology highly similar to that
of yeast, which emerges from a null-model proposed
earlier~\citep{Balcan_etal2007}.

The Boolean regulation dynamics on these networks is investigated by means of
a \textit{synchronous} and \textit{deterministic} update of the network state
as follows: Each node (gene) $i$ has a state $\sigma_i(t)$ at a particular
time $t$ where $\sigma_i(t)$ is either 1 (on) or 0 (off). The network state
$S(t)$ is the set of individual node states: $S(t) = \lbrace
\sigma_1(t),\sigma_2(t),..,\sigma_N(t) \rbrace$.  $\sigma_i(t+1)$ is
determined by the Boolean function $B_i$ assigned to $i$, which is a function
of the states of the neighbor nodes connected to $i$ by incoming edges.  We
used four types of random function classes found in the literature as described
below.

\begin{figure}[!ht] 
 \centering 
  \includegraphics[width=0.9\columnwidth,clip]{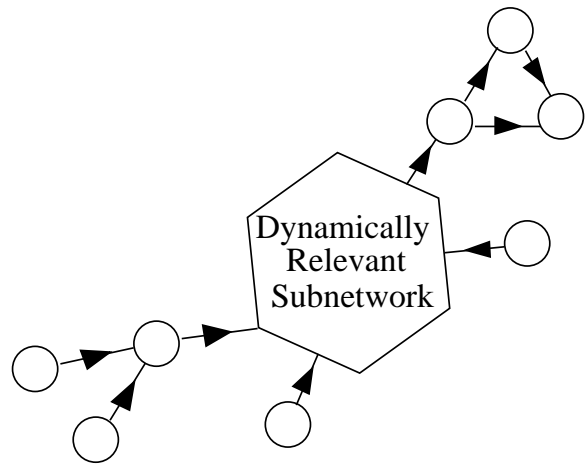}
  \caption[Dynamically Relevant Subnetwork]{The \textit{dynamically relevant
      (sub)network} obtained after recursively pruning the (round) nodes with either zero out-degree or zero in-degree.}
 \label{DynamicallyRelSubnetwork}
\end{figure}

\subparagraph{1- Simple Random Function, \textbf{RF}:}

The rule table is constructed by setting the output for each input combination
to 1 with probabilty $p$ and 0 otherwise, independent of the input.

\subparagraph{2- Canalizing Random Function, \textbf{CF}:}

A subclass of the RFs, that has at least one canalizing input variable whose
canalizing value determines the
output~\citep{Kauffman_Origins_of_Order,Kauffman_Nested}.
\begin{equation}\label{CF}
\begin{split}
B_i(\sigma_{i,1},..,\sigma_{i,j},..,\sigma_{i,k_{in}}) = \\
\left\{ \begin{array}{ll}
 s_{i,j} & \sigma_{i,j}=s_j\\
B_i(\sigma_{i,1},..,\overline{s_j},..,\sigma_{i,k_{in}}) & \sigma_{i,j} \neq s_j
\end{array} \right.
\end{split}
\end{equation}
where j{\it th} in-neighbor is the canalizing node with $s_j$ as the canalizing
value and $s_i$ as the canalizing output. Again, the output is determined
through the parameter $p$. When the canalization condition is not
satisfied, $B_i(\sigma_{i,1},..,\overline{s_j},..,\sigma_{i,k_{in}})$ in
Exps.~\ref{CF} is considered to be a RF.

\subparagraph{3- Nested Canalizing Random Function, \textbf{NCF}:}

\textit{Nested Canalizing} or \textit{Hierarchically Canalizing} functions are
believed to better model gene regulation in biological
systems~\citep{Kauffman_Nested}.  They form a subclass of CFs, where one defines
a canalizing order to the input nodes and the output is determined by the
first node in its canalizing value:
\begin{equation}\label{NCF}
\begin{split}
B_i(\sigma_{i,1},..,\sigma_{i,j},..,\sigma_{i,k_{in}}) = \\ 
\left\{ \begin{array}{ll}
s_{i,1} & \sigma_{i,1}=s_1\\
s_{i,2} & \sigma_{i,1} \neq s_1 \wedge \sigma_{i,2}=s_2\\
... & ...\\
s_{i,j} & \sigma_{i,1} \neq s_1 \wedge \sigma_{i,2} \neq s_2 \wedge ... \wedge \sigma_{i,j}=s_j\\
... & ...\\
s_{i,k_{in}} & \sigma_{i,1} \neq s_1 \wedge \sigma_{i,2} \neq s_2 \wedge ... \wedge \sigma_{i,k_{in}}=s_{k_{in}}\\
\overline{s_{i,k_{in}}} & \sigma_{i,1} \neq s_1 \wedge \sigma_{i,2} \neq s_2 \wedge ... \wedge \sigma_{i,k_{in}} \neq s_{k_{in}}
\end{array} \right.
\end{split}
\end{equation}
We modify the original definition in \citep{Kauffman_Nested}, for the sake of an
unbiased comparison with the other cases, by determining the outputs $\{s_i\}$
with the parameter $p$ as before.

\subparagraph{4- Special Subclasses of Nested Canalizing Random Function, \textbf{SNCF}:}

Following Nikolejewa \textit{et al.}, one can represent the NCFs above in
 a ``minimal logical expression''~\citep{Nikolajewa_SpecTypeRuletable}:
\begin{equation}
\label{SNCF}
\begin{array}{ll}
\sigma_i&=B_i(\sigma_{i,1}, \sigma_{i,2}, ..., \sigma_{i,k_{in}-1}, \sigma_{i,k_{in}})\\
 &=\sigma_{i,1}^\Theta \bigodot (\sigma_{i,2}^\Theta \bigodot (...\bigodot (\sigma_{i,k_{in}-1}^\Theta \bigodot \sigma_{i,k_{in}}^\Theta)...))
\end{array}
\end{equation}
where $\bigodot$ represents either AND or OR logical function, i.e.
$\bigodot\in\lbrace\wedge,\vee\rbrace$ and $\sigma^\Theta$ stands for a
possible negation of $\sigma$, i.e.
$\sigma^\Theta\in\lbrace\sigma,\overline{\sigma}\rbrace$. Upon investigation
of Harris \textit{et al.} data \citep{Harris_Functions}\footnote{private communication} they found that gene regulatory rules
are mainly governed by two subclasses of NCF:
\begin{equation}
\label{SNCF60}
\sigma_{i,1}^\Theta \wedge (\sigma_{i,2}^\Theta \wedge (...\wedge (\sigma_{i,k_{in}-1}^\Theta \wedge \sigma_{i,k_{in}}^\Theta)...))
\end{equation}
and 
\begin{equation}
\label{SNCF30}
\sigma_{i,1}^\Theta \wedge (\sigma_{i,2}^\Theta \wedge (...\wedge (\sigma_{i,k_{in}-1}^\Theta \vee \sigma_{i,k_{in}}^\Theta)...))
\end{equation}
with $66.39\%$ and $29.41\%$ probability of occurrence, respectively. For
these two functions, $p$ is not a free parameter and depends on the topology.

\begin{figure*}[ht] 
 \begin{center}
  \includegraphics[width=1.2\columnwidth, angle=-90]{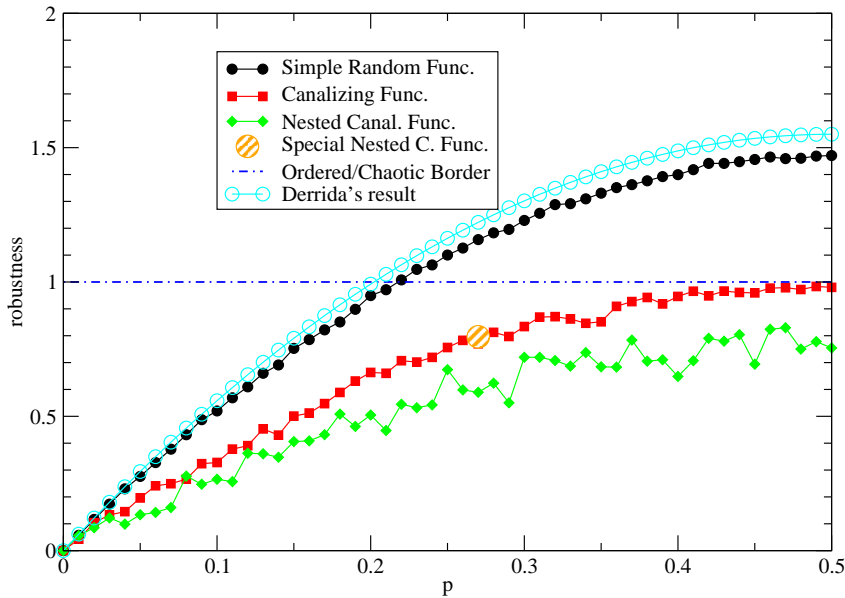}
 \end{center}
  \caption{Robustness of yeast's GRN for all types of functions. For each $p$
    value, robustness was computed with averaging over $1000$ random initial
    conditions of each $10$ realization. Also, the Derrida's Exp.,
    $s=2p(1-p)\langle k_{in} \rangle$ was drawn.}
 \label{Yeast_Robustness} 
\end{figure*}

Once the network topology and the functions are fixed, the Boolean dynamics is
characterized by a set of limit cycles which are the attractors of the
dynamics reached from different initial conditions. Since these are the
regions of the state space where the dynamics converges to, one expects them
to be biologically relevant. For example, they have been
associated with different phenotypes of the plant {\it Arabidopsis
  thaliana}~\citep{Mendoza_etal_Arabidopsis} when the involved genes are those
that take part in cell differentiation. We have investigated and compared the
number, cycle-length, transient length, and the basin of attraction of the
attractors in each case. These results will be presented
elsewhere~\citep{Tugrul_Alkan_GRN}. Here, we deal with another dynamical
property, the robustness of the attractors to perturbation.  We use the
following arguments to measure the robustness as given by
Aldana~\citep{Aldana_BooleanNetPLtopology2003}. Consider two copies of a network
at states $S(t)$ and ${S}^{'}(t)$. Their \textit{Hamming Distance} $HD(t)$
is the number of nodes that differ between the two:
\begin{eqnarray}
\label{HammingDistance}
HD(t)&=&\sum_{i=1}^{N}{\mid\sigma_{i}(t)-{\sigma_{i}^{'}(t)}\mid}\ .
\end{eqnarray}
Let $x(t)\equiv1-\frac{HD(t)}{N}$, where $N$ is the number of the nodes in the network. One defines the robustness $s$ of
the network as
\begin{equation}
\label{sensivity_robustness}
s=\lim_{x\rightarrow1^-, t\rightarrow \infty}\frac{dx(t+1)}{dx(t)}\ .
\end{equation}
The system is robust against perturbations (ordered) if $s<1$, whereas it is
highly sensitive (chaotic) otherwise. It was suggested by
Kauffman~\citep{Kauffman_Origins_of_Order} that the genetic regulatory networks
function at the edge of chaos, where $s\simeq 1$. The quantity $s$ can be
estimated analytically for the RF case under an annealed approximation, both
for random~\citep{Derrida_Pomeau} and power-law
networks~\citep{Aldana_BooleanNetPLtopology2003}. Derrida's result for random
networks is
\begin{equation}
\label{DerridaFormula}
s=2p(1-p)\langle k_{in} \rangle\ ,
\end{equation}
where $p$ is, again, the unbiased probability that a binary function
assigned to a node returns $1$. Note that, by symmetry, $s(p) =
s(1-p)$. We measure $s$ numerically as a function of $p\in [0,0.5]$, by
examining the deviation of the two copies which are initially only slightly
perturbed. For a network with $N$ nodes, the deviation is measured within a
time window of $2N$ steps.

\begin{figure*}[ht] 
  \centering
  \includegraphics[width=1.2\columnwidth,angle=-90]{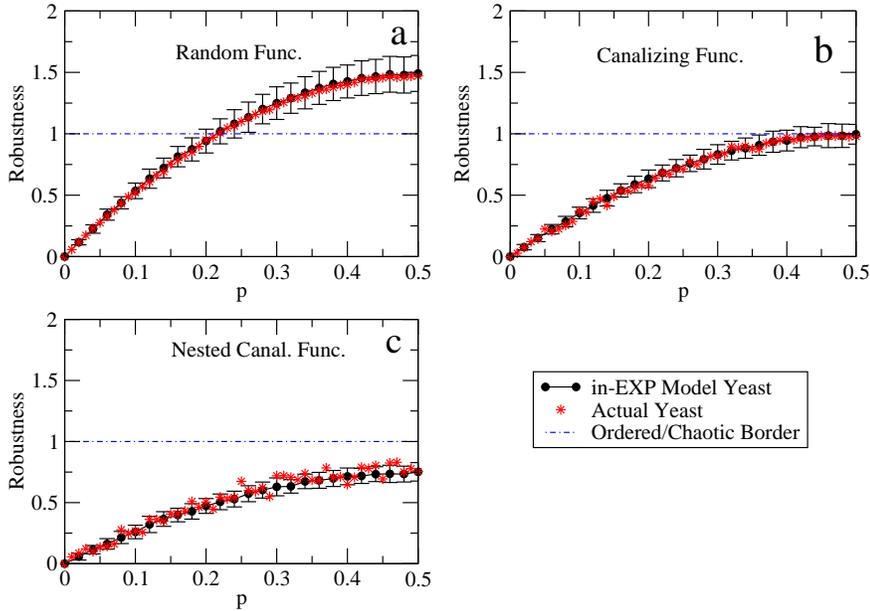}
  \caption{Robustness of in-EXP model networks with RF, CF and NCF type
    regulatory functions. The SNCF function results in $s=0.77 \pm 0.05$,
    whereas for the same $p$ value yeast network has $s=0.78$.}
 \label{EXPModelYEAST_Robustness} 
\end{figure*}

As long as one is interested in the network characteristics such as attractor
statistics or robustness, simulating the dynamics of the whole network is
extremely inefficient. The reason is that, given the topology, some of the
nodes make no contribution to such statistics~\citep{Socolar_Kauffman}. For
example, a node with zero in-degree remains at a fix state all times.
Similarly, a node with zero out-degree simply follows the input and does not
give any feedback. Same statements apply to those nodes which lose all
incoming or outgoing edges after a round of pruning such nodes. Therefore, we
focus on the {\it dynamically relevant (sub)network} (DRN) which is found by
recursively pruning all the nodes with zero in-degree or zero out-degree (see
Fig.~\ref{DynamicallyRelSubnetwork}). This subnetwork is typically much
smaller than the original, allowing one to run time-efficient simulations.

We find that the in-degree distribution of the DRN for the yeast regulatory
network is exponential with an exponent $\alpha = 0.38$, similar to the full yeast network. In addition, we
generated two ensembles of $100$ model networks for comparison. The first
ensemble is a set of randomly connected networks of the same size $N=82$ and
the same exponent in-degree distribution as the yeast DRN, so named in-EXP model. The second ensemble is
generated by using a recently proposed null-model by Balcan {\it et
  al.}~\citep{Balcan_etal2007} which successfully reproduces many topological
features of the yeast's GRN. An interesting observation is that, the second
ensemble which preserves a number of topological signatures found in yeast,
yielded dynamically relevant subnetworks with an average size of $36 \pm 15$,
i.e., significanly smaller than that of the yeast DRN.

\paragraph{RESULTS}

\begin{figure*}[ht] 
 \centering 
  \includegraphics[width=1.2\columnwidth,angle=-90]{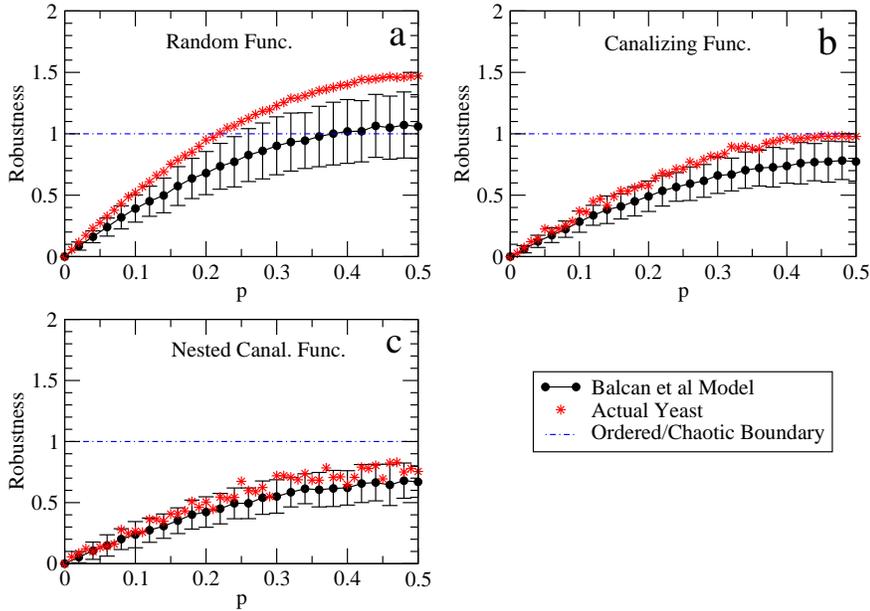}
  \caption{Robustness of Balcan \textit{et al.} model networks with RF, CF and
    NCF type regulatory functions. The SNCF function results in $s=0.83 \pm
    0.08$, whereas for the same $p$ value yeast network has $s=0.78$.}
 \label{Balcan_Robustness} 
\end{figure*}

On the GRN of the yeast, we calculated the robustness of the network dynamics
for each function type discussed above as a function of $p$. We chose $p\in
\lbrace 0.00, 0.01,..,0.50\rbrace$, except for SNCF case, where the value of
$p$ is fixed by the network topology to the value $p=0.27$. For each case, we
performed statistics over $10$ independent function assignments and ran the
dynamics for $2N$ time steps starting from $1000$ different initial conditions
(and their perturbations, in parallel). In all cases, the lengths of the
simulations were sufficient for the system to reach an attractor.
Fig.(\ref{Yeast_Robustness}) shows the average robustness found for each
function type. We find that when a random RF function is associated with each
gene's transcription, the systems switches from an ordered phase to a chaotic
phase around $p = 0.22$, consistent with Derrida's analytical result in
Eq.(\ref{DerridaFormula}). CFs always result in an ordered system, except when
$p=0.5$, where the yeast's GRN appears to be at the edge of chaos. NCF and
SNCFs which have been suggested to better represent gene regulation dynamics
are strictly ordered for all $p$ values.

For comparison, we repeated the same analysis on in-EXP and Balcan \textit{et al.} models. 100 different networks were created from each set in order to
reduce fluctuations due to structural deviations from sample to sample. The
average robustness obtained for in-EXP and Balcan {\it et al.} model are
compared with the corresponding data obtained from the yeast's GRN in
Fig.(\ref{EXPModelYEAST_Robustness}) and Fig.(\ref{Balcan_Robustness}),
respectively. We find that the in-EXP model networks show similar robustness
profiles as the yeast's GRN in all cases, whereas Balcan \textit{et al.}
model, although it globally appears to capture the network
structure~\citep{Balcan_etal2007}, shows a significant deviation
from the yeast in its dynamics.

 \paragraph{DISCUSSION}

We calculated the robustness of the yeast's transcriptional regulatory
network within the framework of Boolean networks and as a function of the
gene activation probability $p$. Under different assumptions on the function
class that governs the regulation process, we find that the network may show
an order-chaos transition with changing $p$, may reach the edge of chaos at
$p=0.5$, or may stay robust for all $p$ values. Our results point to the
fact that, the activation probability by itself is not sufficient to
determine the robustness of the Boolean networks; the functional category of
the update rules also matter. As future experiments more precisely unveil
activation/inhibition relations between genes in the yeast organism, proper
choices for $p$ and the function class shall become apparent. The strong
dependence of the robustness to the function type and $p$ may entail that
both have been optimized throughout evolutionary time scales to their
present-time values. Our findings may then help address the ``edge of chaos''
hypothesis of Kauffman~\citep{Kauffman_Origins_of_Order}.

We furthermore compared our results on the yeast network with those obtained
from two models which produce statistically similar network topologies. We
found to our surprise that among the two models, Balcan {\it et al.} model
which better reproduces a set of global topological features shows
significantly larger deviation from the yeast's network in its robustness.
This discrepancy should stem from certain structural features that are not
captured by the global topological signatures considered in past studies.
One such difference we observe is the much smaller average dynamical core
size of Balcan \textit{et al.} model networks. This and other possible
structural sources for the observed difference in dynamics should be further
investigated.

\section*{Acknowledgments}

 This work was supported by The Scientific and Technological Research Council
 of Turkey Grant TBAG-106T553.


\end{document}